\documentclass[sn-mathphys,Numbered]{sn-jnl}

\usepackage{graphicx}%
\usepackage{multirow}%
\usepackage{amsmath,amssymb,amsfonts}%
\usepackage{amsthm}%
\usepackage{mathrsfs}%
\usepackage[title]{appendix}%
\usepackage{xcolor}%
\usepackage{textcomp}%
\usepackage{manyfoot}%
\usepackage{booktabs}%
\usepackage{algorithm}%
\usepackage{algorithmicx}%
\usepackage{algpseudocode}%
\usepackage{listings}%
\usepackage{longtable}

\begin{document}
\title{Nuclear Isomers at the Extremes of their Properties}

\author[1]{\fnm{Bhoomika} \sur{Maheshwari}}\email{bhoomika.physics@gmail.com}

\author[2,3]{\fnm{Ashok} \sur{Kumar Jain}}\email{ashkumarjain@yahoo.com}

\affil[1]{\orgdiv{Department of Physics, Faculty of Science}, \orgname{University of Zagreb}, \orgaddress{ \city{Zagreb}, \postcode{HR-10000},  \country{Croatia}}}

\affil[2]{\orgdiv{Amity Institute of Nuclear Science and Technolgy}, \orgname{Amity University UP}, \orgaddress{ \city{Noida}, \postcode{201313}, \country{India}}}

\affil[3]{\orgdiv{Department of Physics}, \orgname{Indian Institute of Technology Roorkee}, \orgaddress{ \city{Roorkee}, \postcode{247667}, \country{India}}}

\date{\today}

\abstract{ The longer-lived excited nuclear states, referred as nuclear isomers, exist due to the hindered decays owing to their peculiar nucleonic structural surroundings. Some of these conditions, being exceptionally rare and limited to achieve, elevate certain isomers to the status of extreme and unusual isomers among their kin. For example, the $E5$ coupling of single-particle orbitals is rare and so are $E5$ decaying isomers. This review delves into some of such remarkable isomers scattered across the nuclear landscape while highlighting the possibilities to find more of them. Unique properties of some of them, harbor the potential for transformative applications in medicine and energy. An exciting example is that of the lowest energy isomer known so far in $^{229}$Th, which may help realize the dream of an ultra-precise nuclear clock in the coming decade. These isomers also offer an insight into the extremes of nuclear structure associated with them, which leads to their unusual status in energy, half-life, spin etc. The review attempts to highlight isomers with high-multipolarities, high-spins, high-energies, longest half-lives, extremely low energy, etc. A lack of theoretical understanding of the decay rates, half-lives and moments of these isomers is also pointed out.}

\keywords{Very high-multipolarity/high-spin/high-energy/long-lived isomers, Extremely low energy isomers, Proton-decaying isomers, $\beta$-decaying isomers}

\maketitle
\section{Introduction}

Nuclear isomers are the nuclear meta-stable states which occur throughout the nuclear landscape~\cite{Book}. The Second Edition of the Atlas of Nuclear Isomers~\cite{Garg2023}, which is an updated and revised version of the 2015 Atlas of Nuclear Isomers~\cite{Jain2015}, presents about 2600 isomers with a half-life $\ge 10$ nanoseconds. The number will obviously increase if we lower the half-life limit to 1 nanosecond. The very first theoretical interpretation of most of the isomeric states was given by Weizsacker~\cite{Weizsacker1936} by invoking the large change in angular momentum during the decay process to be instrumental in giving rise to large life-times; these isomers are now known as spin isomers. Isomers and their correlation with shell structure is now well established in spherical as well as deformed nuclei~\cite{Feenberg1949,Bohr}. Besides the spin isomers, we also have other physical mechanisms which may lead to isomers of other types~\cite{Walker1999}. Most prominent of these are, shape isomers~\cite{Moller2012}, fission isomers~\cite{Singh2002}, K-isomers~\cite{Kondev2015,Walker2016}, and seniority isomers~\cite{Isacker2011,Maheshwari2022,Book}. It is entirely possible for more than one mechanisms to be responsible for the formation of a given isomeric state~\cite{Walker2005,Book}. With the advent of several radioactive ion beam facilities, it should become possible to find and investigate isomers in nuclei at the extremes of the nuclear chart. Therefore, an exciting era of new discoveries in isomers awaits us. 

This article has a very limited scope of pointing out some of the extreme and unusual examples of isomers which have some unique property in terms of spin, excitation energy, $N/Z$ ratio, decay transition etc. Many of these isomers could be very useful in potential applications in medicine, energy, astrophysics, ultra-precision clocks, gamma-ray lasers~\cite{Walker2020,Book} etc. In doing so, we have largely relied upon the Atlas of Nuclear Isomers - Second Edition \cite{Garg2023}. A complete understanding of the properties of these unusual isomers and the expected new findings in the drip-line regions, is going to pose a major challenge to the existing nuclear theories.

The paper is organized as follows. In the following sections, we discuss selected examples of high-energy isomers, high-multipolarity isomers, extremely low energy isomers, very high-spin isomers, very long-lived isomers, proton-decaying isomers, and $\beta-$decaying isomers. The highest multi-quasi-particle isomer in $^{175}$Hf is also briefly discussed. The futuristic isomer application in dark matter research is highlighted. The last section concludes the paper.

\section{High energy isomers}

In general, one does not expect many isomers to occur at high excitation energies for the simple reason that such states might always find a faster decay path due to the possibility of decay to one of the many lower-lying states. Even then, several isomers have been observed at very high excitation energies. We present few interesting examples here.

The ${28}^-$ isomer in $^{208}$Pb with an excitation energy of 13.67 MeV and 60 $ns$ half-life is probably the highest lying isomer known so far. It has an exceptionally large spin as well as high excitation energy. It is rather interesting that $^{208}$Pb is host to almost 8 high energy isomeric states, many having a half-life between 1 ns to 10 ns~\cite{Book}. This is largely due to the double magic nature of this nucleus, which imparts it a very large shell gap and a proximity to high-j single particle levels like $i_{13/2}$, $j_{15/2}$, $g_{9/2}$ orbitals near the Fermi energy. Interpretation of such isomeric states provides a very stringent test of the modern shell model calculations.
    
The $31$ year ${16}^+$ isomer in $^{178}$Hf lies at an excitation energy of $2.4$ MeV, something very unusual for an isomer which is so long-lived. It supports an interesting rotational band known up to ${22}^+$ with a nearly constant moment of inertia. This nucleus became a crucial landmark in the development of the rotational model in early days~\cite{Bohr}. Interest in this isomer arises from its capacity as an energy storage device as well the possibility to obtain a high energy gamma ray laser~\cite{Walker2005}. Due to its large half-life, $^{178}$Hf isomer may also be used as a target in nuclear reactions, or an isomer beam may also be created to carry out some unique nuclear reactions and products which are otherwise inaccessible~\cite{Dracoulis2016,Walker2017}.

Yrast traps having high excitation energy can also arise in the neutron-rich light mass even-even nuclei in and around the ``island of inversion", where the ordering of the single particle orbitals may change due to the changing shell gaps. An interesting example of such an isomer has recently been observed in $^{32}$Si $(Z=14, N=18)$. Williams \textit{et al.}~\cite{Williams2023} reported a new measurement of the $5^-$, 5505.2(2) keV isomer decaying by an unusually high energy transition of 3562.84 (14) keV. Considering the large mean life of 46.9(5) ns, an $E3$ nature was assigned to the transition. These measurements also modify the earlier values reported in the Atlas \cite{Garg2023}. The $4^+$ state (5881 keV) now lies higher than the $5^-$ isomer. The isomer is claimed to be located in a yrast trap, which forms due to the relevant yrast states containing different configurations sensitive to proton and/or neutron cross-shell excitation across the $Z=14$ subshell gap and the $N=20$ shell gap. The $5^-$ state acquires a dominance of the $f_{7/2}$ neutron orbital occupancy. This is a rare example of a spin isomer in an even-even sd-shell nucleus, decaying by a very high energy gamma ray.

\section{High multipolarity isomers}

\begin{figure*}[!htb]
\centering
\begin{minipage}{0.55\textwidth}
    \centering
    \includegraphics[width=\textwidth]{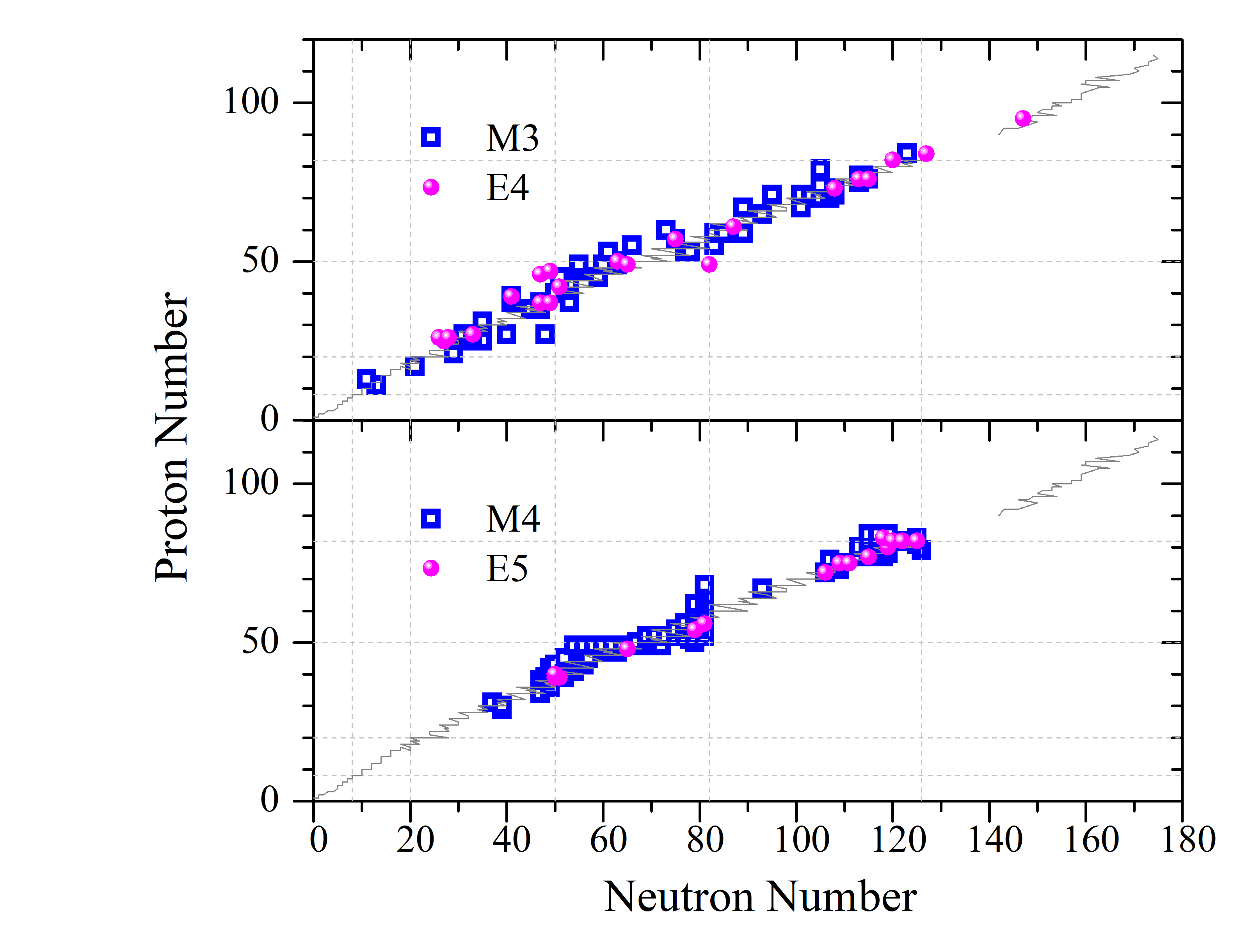}  
\end{minipage}
\begin{minipage}{0.3\textwidth}
   \centering
   \includegraphics[width=\textwidth]{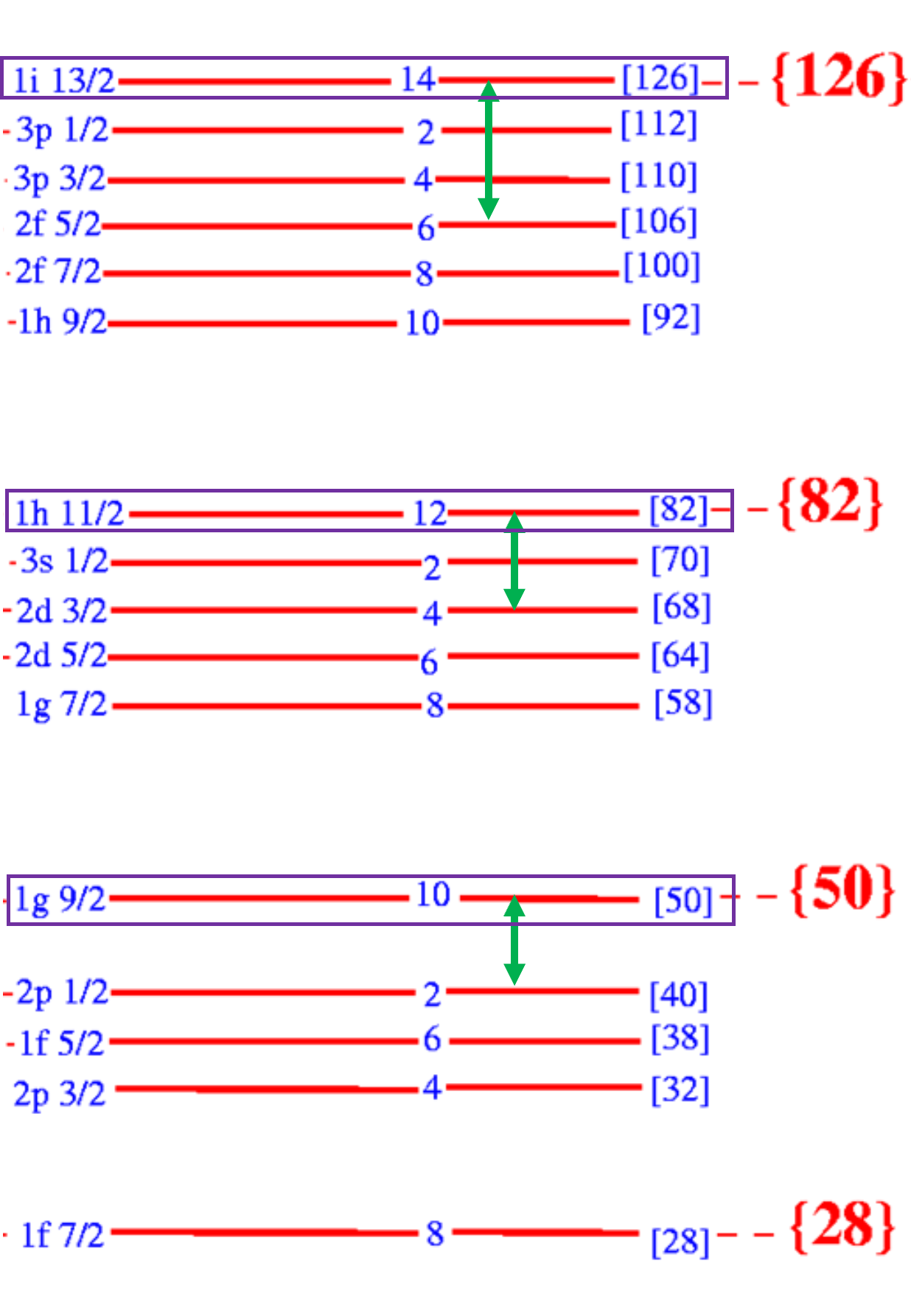} 
\end{minipage}
\caption{\label{fig:multipolarity}(Color online) $\textit{Left:}$ Occurrence of high-multipolarity isomers shown on a $(N,Z)$ chart. Also, shown are the magic number lines. Data taken from the Atlas of Nuclear Isomers-Second Edition~\cite{Garg2023}. The upper panel refers to the $M3,E4$ decaying isomers while the lower panel shows the $M4,E5$ decaying isomers. $\textit{Right:}$ A part of schematic shell model level scheme above nucleon number 28. The intruder orbitals are highlighted which may couple by $M4$ coupling to nearby orbitals. This can be correlated with the occurrence of $M4$ isomeric clusters near magic number lines in the $(N,Z)$ chart. } 
\end{figure*}

\subsection{M3 and M4 Isomers}

By high multipolarity isomers, we generally mean isomers decaying by Isomeric Transition (IT) having $L\ge3$. A number of low lying nuclear isomers decaying by $M3$ and $M4$ transitions have been observed in light mass nuclei; their number is in fact much more than those decaying by $E4$ or, $E5$ transitions, respectively. We hereby discuss the low lying isomers decaying by magnetic octupole $(M3)$ transitions. We can make two quick observations about these isomers. One, nearly all such isomers have been observed in odd-odd nuclei only. Two, most of them are located in two pockets, namely Z=11-27 (8 isomers) and Z=35-53 (14 isomers) proton numbers in the light mass region, although few cases are observed in higher mass regions also. Their half-lives are in general quite large, ranging from milliseconds to hours. 
 
The longest lived $M3$ isomer is observed in $^{58}$Co at a very low excitation of 24.95(6) keV and spin $5^+$, having a half-life of 8.853(23) hours. Another very low lying $M3$ isomer has been found in $^{102}$Ag, which is a $2^+$ isomer at 9.40(7) keV having a half-life of 7.7(5) min. The lowest lying $M3$ isomer has been observed in the odd-odd $^{142}$Pr, which has an excitation energy of merely 3.694(3) keV, half-life of 14.6(5) min and spin $5^-$. Generally, all pure $M3$ decaying isomers have an excitation energy much lower than 500 keV, one of the highest being $^{38}$Cl at 671.365(8) keV from Atlas. Since, there are no high-j orbitals in the light mass region, a coupling of odd neutron and odd proton in an odd-odd nucleus can give rise to the formation of a closely spaced pair of a high and a low-angular momentum state; this could be the reason for the formation of a spin trap in the light mass odd-odd nuclei. Low energy and high multipolarity of the transition ($M3$) contribute to the large half-life seen in these isomers. 

The $M3$ decaying isomers are scattered all along the line of beta-stability of normal nuclides as shown in the left of Fig.~\ref{fig:multipolarity}. In contrast, the parity changing $M4$ decaying isomers are observed to be clustered near the magic numbers $Z=N=50$, and $Z=N=82$. This can be understood by the location of intruder orbitals near Fermi surface and their direct $\Delta L=4$ coupling with the available orbitals in the valence space, as shown by the schematic shell model level scheme in the right side of Fig.~\ref{fig:multipolarity}. All the $M4$ decaying isomers are low lying in energy ($\leq 500$ keV) with few exceptions. All have long half-lives from seconds, minutes, hours, days, to years. An exception is the isomer in $^{206}$Tl, which lies at a high excitation energy of 2643.10(18) keV, has a spin (${12}^-$) and half-life of 3.74(9) min. 

The number of pure $E4$ decaying isomers is relatively very small, and are scattered around the line of stability. We do not see any specific pattern in the occurrence of $E4$ decaying isomers. Only a few cases do exhibit a mixed $M3+E4$ transition.

\subsection{E5 isomers}
The highest multipolarity isomers observed so far are the $E5$ decaying isomers. We present some rare examples of the $E5$ isomers in the following.

A very long-lived isomer is the $(8^+)$ isomer in the odd-odd $^{186}$Re $(Z=75,N=111)$ with $2 \times {10}^5$ $y$ half-life at 148.2 keV energy. This isomer has an $E5$ decay as well as $\Delta K=5$ transition, where $K$ is the projection of total angular momentum along the symmetry axis. It may be a combination of spin and K-isomerism. 

The ${11}^-$ isomer in $^{192}$Ir $(Z=77,N=115)$ has a 241 $y$ half-life and 168.14 keV excitation energy. This is again an odd-odd nuclide with an hindrance due to large change in spin as well as low-energy transition, both contributing to the long half-life. 

The ${16}^+$ isomer in $^{178}$Hf $(Z=72,N=106)$ exists with a 31 $y$ half-life at 2.44 MeV energy and decays by an M4 as well as E5 transition. This case has already been discussed in previous section as an example of a high-energy long-lived isomer. The region around $^{178}$Hf is known for yrast isomers, where the contribution of collective rotation to the angular momentum is minimal and spin-alignment of involved orbitals is maximal. This is possible when the Fermi surface is surrounded by high-j nucleon orbitals. The isomer was assigned a four-quasiparticle configuration of $\pi 7/2[404]9/2[514] \otimes \nu 7/2[514] 9/2[624]$ using projected shell model~\cite{Sun2004}. However, a theoretical understanding of the measured isomer decay probability is still missing. This is true for many deformed isomers, where efforts are very much required to reproduce the known decay properties quantitatively. 

The ${11/2}^-$ isomer in $^{113}$Cd $(Z=48,N=65)$ has 13.89 $y$ half-life and 263.54 keV excitation energy.
This isomer decays with $\%$ IT $=0.14$ and $\% \beta^{-} =99.86$. Though the main isomer decay branch is $\beta^-$decay but the isomer also has a small $E5$ gamma branch to the respective ground state ${1/2}^+$. The isomeric configuration is mainly dominated by $h_{11/2}$ orbital though the mixing of nearby low-j orbitals is found to be crucial to explain the isomeric Q-moments~\cite{Maheshwari2019}.

The region around $^{208}$Pb has the potential to find new high multipolarity isomers due to the vicinity of many high-j orbitals lying alongside low-j orbitals as shown in Fig.~\ref{fig:multipolarity}. When the $i_{13/2}$ orbital from the 82-126 nucleonic space couples with the $f_{5/2}$ orbital, it has the potential to generate $M4$ and $E5$ transitions, making it a plausible region for the presence of such high-multipolarity isomers. For example, the $9^-$ isomers in $^{202,204}$Pb isotopes undergo decay through $E5$ transitions to lower-energy $4^+$ states. Consequently, one might anticipate the existence of a similar $9^-$ isomer in $^{206}$Pb. Nevertheless, the specific location of another isomer, the $7^-$ isomer, already identified in $^{202,204,206}$Pb, influences the decay pathway of the $9^-$ state in $^{206}$Pb which is connected by $E2$ decay to the lower-lying $7^-$ isomer in this nuclide. Yet, we can expect more of large multipolarity isomers in this mass region.

\section {Extremely Low Energy (ELE) isomers} 
    
The half-life of an isomeric state is inversely proportional to the decay energy and the multipolarity. It is, therefore, possible to have isomers only because of extremely low excitation energy even when the multipolarity of the decaying transition is small.  These isomers should not have large spin. Therefore, "no hindrance due to angular momentum" makes these isomers very unique. The examples, which have an excitation energy less than 2 keV, are listed in Table~\ref{tab:eletable}, except for $^{142}$Pr which has an excitation of 3.694 keV.
\\
\begin{center}
\begin{longtable}{ c  c  c  c c}
\caption{\label{tab:eletable} Extremely Low Energy (ELE) Isomers}
\\
\hline \\
$A^X$ & Energy (keV) & $J^\pi$ & Multipolarity & Half-life \\
\hline    \\
$^{110}$Ag & 1.112(16) & $2^-$ & $E1$	& 660(40) $ns$ \\
$^{141}$Sm & 1.58(4) & ${3/2}^+$ &	& 	\\
$^{193}$Pt & 1.642 (2) & ${3/2}^-$ &  $M1$ & 9.7(3) $ns$ \\
$^{201}$Hg & 1.5648(10) & ${1/2}^-$ &	$M1+E2$ &	81(5) $ns$ \\
$^{227}$Ra & 1.733(9) & ${5/2}^+$ & & \\
$^{229}$Th	& 0.008338(24) & ${3/2}^+$ & $M1$ & 7(1) $\mu s$  \\
$^{235}$U &	0.076737(18) & ${1/2}^+$ &	$E3$ & 26.5(10) $min$ \\
$^{142}$Pr & 3.694(3) & $5^-$ & $M3$& 14.6(5) $min$ \\
\hline
\end{longtable}
\end{center}

\textbf{$^{229}$Th:} The ${3/2}^+$ isomer in $^{229}$Th $(Z=90, N=139)$ with an excitation energy of 8.338(24) eV is the lowest energy isomer known so far. This energy corresponds to the vacuum ultra-violet region of light. A half-life of 7.0(10) $\mu s$ for a neutral atom was measured from conversion electron measurement. 
As pointed out in the Atlas~\cite{Garg2023}, the chemical environment can change the half-life significantly. For doubly-charged ions, the half-life becomes $> 60$ $s$, as electron conversion process gets shut off. Its radiative decay was seen for the first time by Kraemer \textit{et al.}~\cite{Kraemer2023} and the radiative half-life measured to be greater than 670 (102) sec. The knowledge of a precise gamma decay energy should make it possible to find a laser for direct excitation of the isomer and its use in an optical nuclear clock. It is probably the most famous nuclear isomer of the present times for potential applications. 

The structure of $^{229}$Th was briefly discussed by Jain \textit{et al.}~\cite{Jain1990} in 1990, who considered this nucleus to be sitting at the boundary of octupole and normal quadrupole deformation. The octupole deformation decreases as the neutron number increases and $N=139$ sits right at the border of the transition. As pointed out, the ${5/2}^+$ ground state corresponds more nearly to the $5/2[633]$ Nilsson orbital. However, just above the ground state, at 8 eV, a ${3/2}^+$ state is observed which is formed from a parity-mixed orbital ${3/2}^+[0.3, 0.3]$, (where the numbers in square parentheses represent the expectation values of the single particle operators $s_z$ and $\pi$ for an reflection-asymmetric potential having an octupole component). Thus we have the coexistence of a normal Nilsson orbital and a parity-mixed single-particle orbital within 0.01 keV of each other. Indeed, the ${5/2}^+$ ground state has a normal octupole vibrational state ${5/2}^-$ at 512 keV, and the ${3/2}^+$ isomeric state has an octupole parity doublet ${3/2}^-$ at 164 keV. This coexistence of normal and octupole states may be due to the polarization of the soft core by the odd particle in different orbitals towards smaller or larger octupole deformation. A more detailed theoretical investigation along these lines would be highly interesting.

\begin{figure*}[!htb]
\centering
\includegraphics[width=0.80\textwidth,height=0.5\textwidth]{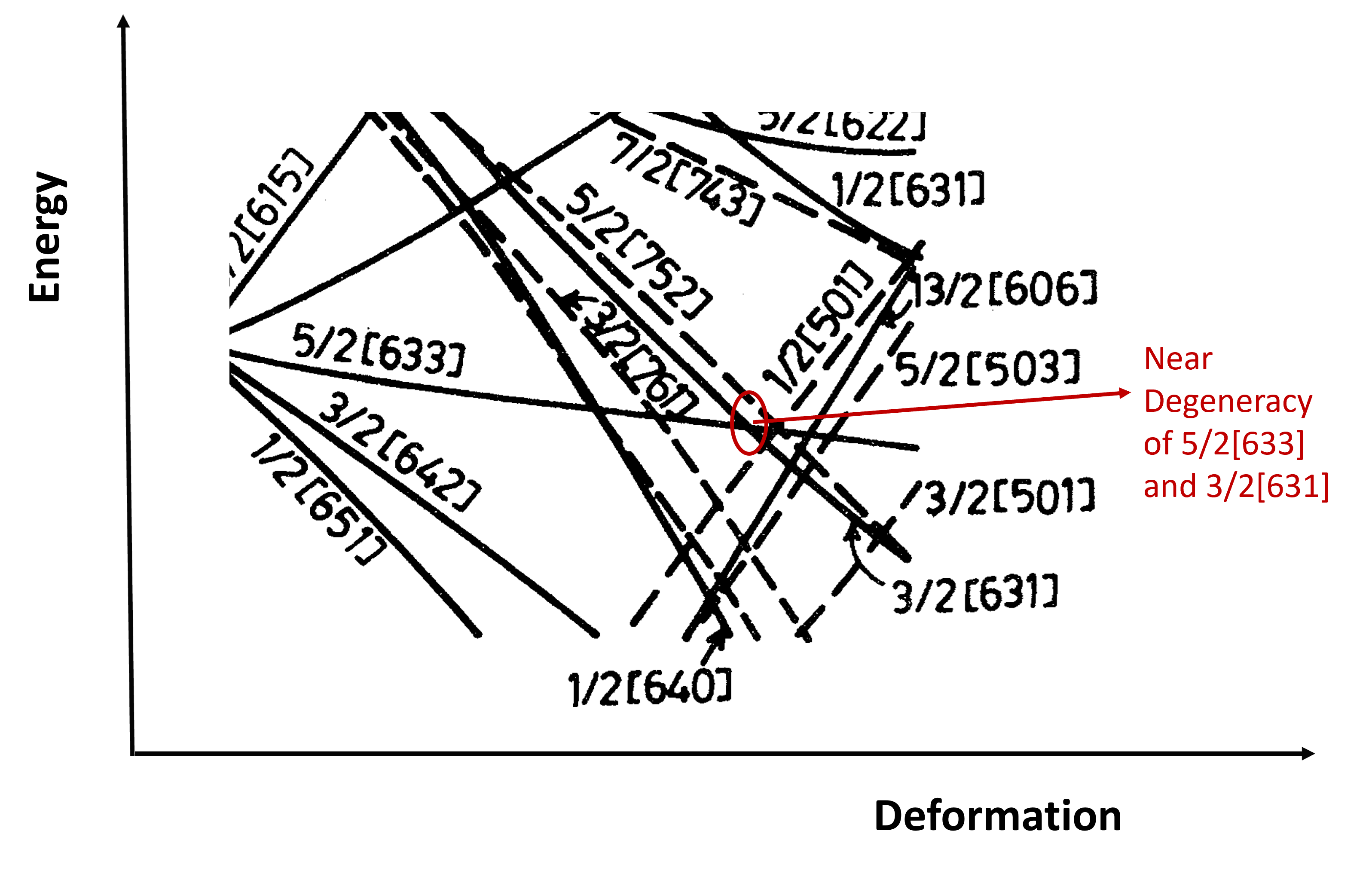}
\caption{\label{fig:nilsson}(Color online) A part of the Nilsson scheme~\cite{Jain1990} where one can easily see the near degeneracy of 5/2[633] and 3/2[631] orbitals for the 139th neutron near the Fermi surface in $^{229}$Th.} 
\end{figure*}
 
\textbf{$^{235}$U:} The ${1/2}^+$ isomer in $^{235}$U $(Z=92, N=143)$ having an excitation energy of 76.7 eV is another example of an extremely low energy nuclear isomer. It has a reasonably large half-life of 26.5 $min$, which may be significantly affected by the chemical environment due to the low energy of the conversion electrons. The decay proceeds via an $E3$ transition to the ${7/2}^-$ ground state. Therefore, large multipolarity also has a role to play in the hindrance to the transition. The Nilsson single particle scheme for neutrons by Jain \textit{et al.}~\cite{Jain1990} clearly exhibits the coming together of the 1/2[631] and 7/2[743] neutron orbitals at a deformation ${\epsilon_2}=0.26$. In addition, 5/2[622], and few other neutron orbitals are also coming close to these orbitals, as seen in the Nilsson scheme. Most of these levels are, however, yet to be identified properly. Further, as shown in the empirical systematics of Figures 33 and 35 of Jain \textit{et al.}~\cite{Jain1990}, many nuclides like $^{229}$Pa, $^{223}$Ac, and $^{233}$Th are promising candidates for such ELE isomers. As we go higher and higher in the masses, the single particle levels get denser and possibilities of their close crossings increase, thereby giving rise to ELE isomers.

\textbf{$^{110}$Ag:} The $2^-$ isomer in $^{110}$Ag $(Z=47, N=63)$ decays by an E1 transition to the $1^+$ ground state. It lies just 1.112 keV above the ground state and has 660(40) $ns$ half-life. The two-quasi-particle configuration of $2^-$ is quite different from that of the $1^+$ ground state. The decay is thus hindered by the low energy transition as well as differing configurations. 

\textbf{$^{201}$Hg:} The ${1/2}^-$ isomer in $^{201}$Hg $(Z=80, N=121)$ having 1.5648(10) keV energy with 81(5) $ns$ half-life decays to the ${3/2}^-$ ground state. Since there is no angular momentum hindrance involved, it is a true ELE isomer.

\textbf{$^{193}$Pt:} The ${3/2}^-$ isomer in $^{193}$Pt $(Z=78, N=115)$ having 1.642(2) keV energy with 9.7(3) $ns$ half-life decays to the ${1/2}^-$ ground state. This also appears to be an ELE isomer.

\textbf{$^{142}$Pr:} The $5^-$ isomer at 3.694(3) keV in $^{142}$Pr $(Z=59, N=83)$, having half-life of 14.6(5) min decays by an M3 transition to the $2^-$ ground state. It appears to be a mixture of low energy and spin isomer.

There are two more extremely low lying states where experimental data are incomplete. The nucleus $^{141}$Sm $(Z=62, N=79)$ has a ${3/2}^+$ level at 1.58(4) keV, and the nucleus $^{227}$Ra $(Z=88, N=139)$ has a ${5/2}^+$ level at 1.733(9) keV but no measured half-lives. Further measurements are required to confirm these as isomers.

\section{Very high-spin isomers}

We note that very high-spin isomers are more likely to have a high excitation energy also, because most of them have a multi-quasi-particle configuration. Also, most of these cases have been observed in the trans-lead region (At, Rn and Fr etc.) and may be understood based on particle-vibration coupling model~\cite{Hamamoto1974}. The interaction between the single-particle degrees of freedom and the octupole vibrational mode allows emergence of such higher-spin isomers in nuclei not very far from the closed shell. Hence, most of these cases are known to decay with an $E3$ transition. The recent shell model study of octupole excitations near $^{208}$Pb~\cite{Isacker2022} could be an important basis for understanding these very high-spin and high-energy isomers in trans-lead region. We discuss a few examples below.  

\textbf{$^{213}$Fr:} With five active protons outside the $Z=82$ core, this nuclide has five known isomers having a half-life $T_{1/2}>10$ nanoseconds. The longest-lived among them is the $({65/2}^-)$ isomer which was discovered in 1989 based on the predictions of the particle-vibration model~\cite{Byrne1989}. The $({65/2}^-)$ isomer in $^{213}$Fr (Z=87, N=126) lies at excitation energy of 8.09 MeV having 3.1 $\mu s$ half-life. Experimental identification of such high-spin isomer was very crucial to establish the applicability of the particle-vibration model at high spins. Byrne et al.~\cite{Byrne1989} identified this highest-spin isomer in trans-lead region, and measured its mean life, decay modes, and the g-factor as 0.695 (7) nm. It is an $E3$ decaying isomer with a mixed proton-neutron configuration as $\pi {[h_{9/2}^3 i_{13/2}^2]}_{{45/2}^-} \otimes \nu {[p_{1/2}^{-2} g_{9/2} i_{11/2}]}_{{10}^+}$. 

\textbf{$^{212}$Rn:} With $Z=86, N=126$, $^{212}$Rn exhibits a relatively simple structure comprising a closed neutron core and four valence protons. Surrounded by many high-j orbitals makes it an intriguing candidate for investigating both low- and high-spin isomers. It has 16 isomers known so far and has emerged as an island of many high-spin isomers having spin ${30}^+$, ${34}^-$, ${37}^-$, and ${38}^+$ at very high excitation energies. The existence of such high-spin and high-energy isomers suggests the role of the neutron core excitation from $N=126$ closed shell, once the angular momentum available from the valence protons is exhausted. The low-lying levels are dominated by four-proton configurations in $h_{9/2}, i_{13/2}$ and $f_{7/2}$ orbitals. The maximum positive and negative spin and parity from the four protons in the $h_{9/2}, i_{13/2}$ and $f_{7/2}$ orbitals can be ${20}^+$ and ${21}^-$, respectively. The ${30}^+$ isomer in $^{212}$Rn (Z=86, N=126) was known at 8.57 MeV energy with $154$ $ns$ half-life ~\cite{Horn1977,Horn1979}. In 1990, Dracoulis \textit{et al.}~\cite{Dracoulis1990} identified the next high-spin and high-energy isomer with spin ${33}^-$. The ${34}^-$ isomer sits at 10.61 MeV energy with $\approx 20$ $ns$ half-life.  In 2009, Dracoulis et al.~\cite{Dracoulis2009} again studied this nucleus and reported more high-spin ${({37}^-)}$ and ${({38}^+)}$ isomers at energies $>12$ MeV. This makes the ${({38}^+)}$ isomer at 12.5 MeV energy with 8 ns half-life as the highest-spin isomer in this nuclide so far. 


\textbf{$^{151}$Er:} Another interesting example is the $({67/2}^-)$ isomer in $^{151}$Er (Z=68, N=83) at 10.28 MeV energy with 0.42 $\mu s$ half-life. The high-spin states in this nuclide are expected to arise from the alignment of $h_{11/2}$ protons and $f_{7/2}$, $h_{9/2}$ and $i_{13/2}$ neutrons. The 0.42 $\mu s$ high-spin isomer was first identified by Andre et al in 1990~\cite{Andre1990}. The same group~\cite{Foin2000} did a subsequent study to establish the decay mode and spin-parity assignments by adding the conversion electron measurements. A possible mixed neutron-proton configuration for this high-spin isomer was also suggested as  $\pi {[h_{11/2}^4 d_{3/2}^1 d_{5/2}^{-1}]}_{{20}^+} \otimes \nu {[f_{7/2} h_{9/2} h_{11/2}^{-1}]}_{{27/2}^-}$~\cite{Foin2000}. It may be noted that the high-spin spectroscopy is also known for its neighboring $N=83$ isotonic nuclei, $^{147}$Gd and $^{149}$Dy but without any similar high-spin isomer. This makes $^{151}$Er a unique case in this mass region so far known with $E3$ decaying high-spin and high-energy isomer. Newer experiments are required in this region to reveal if $^{151}$Er is really a unique case. 


It is apparent that a region with many active high-j orbitals along with the possibility of an octupole interaction near closed shells can give rise to many intriguing candidates for high-spin and high-energy isomers. The region around $^{208}$Pb is already well known. One may expect that another ideal region for the high-spin isomers could be around $^{146}$Gd due to the active proton $h_{11/2}$ and neutron $f_{7/2}$, $h_{9/2}$ and $i_{13/2}$ orbitals. More efforts to find high-spin isomers in this mass region may become fruitful.

\section {Highest quasi-particle isomer}

The Hf-W-Os region is known for prolate-deformed multi-quasi-particle high-K isomeric states built on large-$\Omega$ Nilsson orbitals from deformation-aligned high-j
levels near the Fermi surface, where K/$\Omega$ are the projection of total/single-particle angular momentum on symmetry axis for axially-deformed nuclei. The highest number of quasi-particles known in an isomer is the nine-quasiparticle $(57/2^-)$ isomer in $^{175}$Hf lying at 7.455 MeV having a half-life of 22 $ns$~\cite{Gjorup1990}. The isomeric configuration was suggested to be built upon a seven-quasi particle ${45/2}^+$ isomer, further coupled with $\pi 5/2 [402] \otimes \pi 7/2[523]$. This isomer is unique since it does not decay following the available K-allowed gamma routes but hinders the decay by following the 661 keV, $M1$ gamma to spin $55/2$, which is 9-10 times K-hindered transition. 

It may be noted that the high-spin isomers around shell closure (as discussed in section 5) and the high quasi-particle isomers in deformed region arise due to different mechanisms. The isomers based on large number of quasi-particle configurations in deformed nuclei are of special interest as these may provide an insight in the vanishing pairing correlations and approach to the rigid body regime. A widely used theoretical framework of such multi-quasiparticle states was first given by Kiran Jain et al.~\cite{Kiran1995}. However, a more detailed theoretical model is still required to explain all the properties including half-life of these isomers.

\section {Very long-lived isomers}

\begin{figure*}[!htb]
\centering
\includegraphics[width=0.90\textwidth,height=0.65\textwidth]{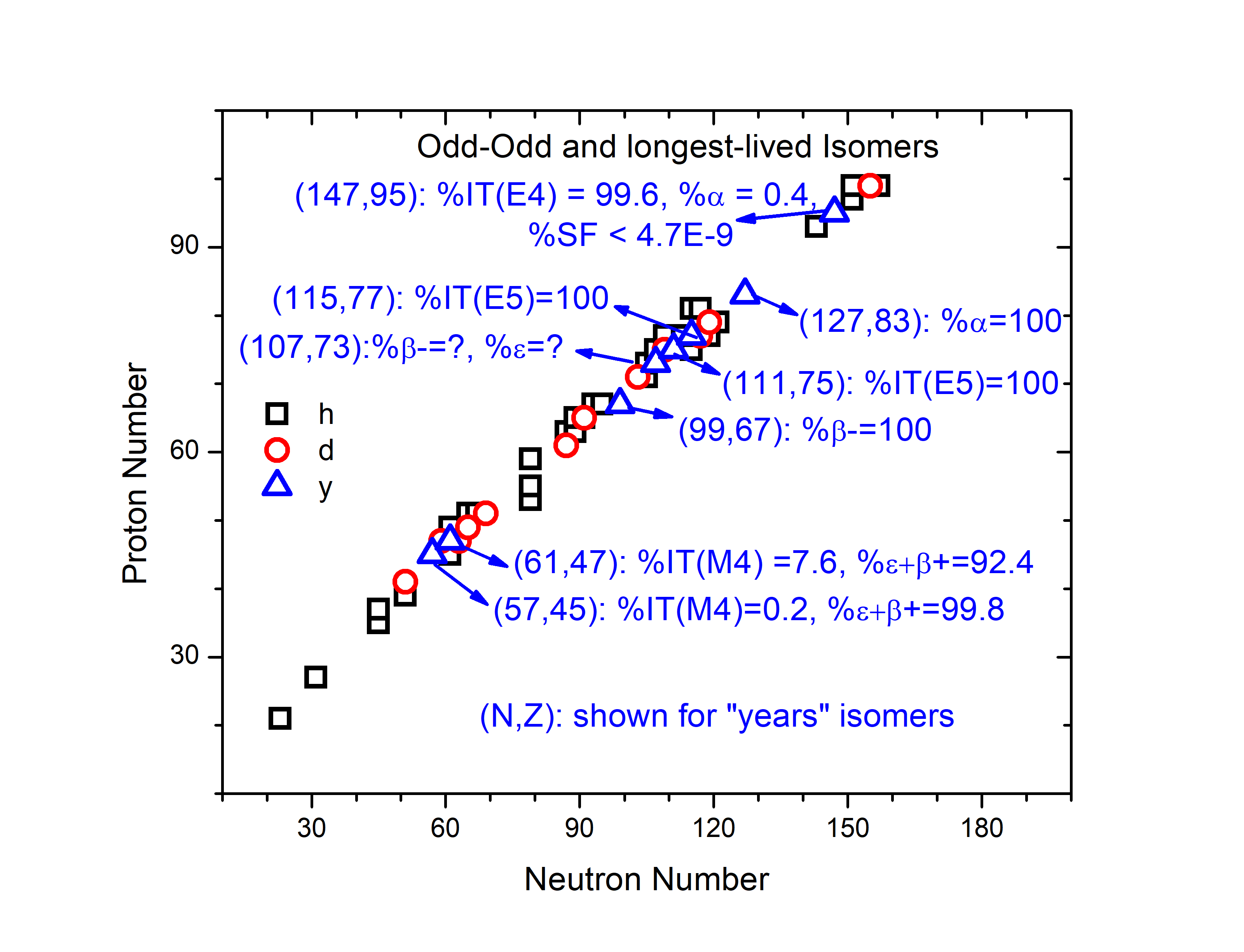}
\caption{\label{fig:longestlived}(Color online) $(N,Z)$ chart of the longest-lived odd-odd isomers with half-lives in hours (h), days (d) and years (y); data taken from the Atlas of Nuclear Isomers-Second Edition~\cite{Garg2023}. The text in blue gives the $(N,Z)$ of the isomers having half-lives in ``years", and their decay modes.} 
\end{figure*}

The odd-odd nuclei are conspicuously known to have largest number of isomers, many of them long-lived. This is largely due to the neutron-proton multiplet arising at low excitation energies.The 2023 Edition of Atlas of Nuclear Isomers~\cite{Garg2023} contains a total of 12 isomers with half-lives in ``years (y)". Eight of the 12 cases belong to the odd-odd nuclei and are shown in Fig.~\ref{fig:longestlived}.  Similarly, 12 of the 39 isomers having half-lives in ``days (d)", and 36 out of the 67 isomers having half-life in ``hours (h)" also belong to odd-odd nuclei. There is, thus, a preponderance of odd-odd nuclei among the long lived isomers also. Many of these long-lived odd-odd isomers, such as $^{108}$Ag, $^{166}$Ho, $^{186}$Re, $^{192}$Ir, and $^{210}$Bi play an important role in the s-process nucleosynthesis.  We present some selected examples of long-lived isomers in the following.

$^{180}$Ta: The $9^-$ isomer at 76.79(55) keV in $^{180}$Ta $(Z=73, N=107)$ has a half-life of $>\ 4.5 \times {10}^{16}$ $y$, the longest lived isomer known so far. In addition, this odd-odd nuclide has 13 known isomers with half-lives $>$ 1 nanosecond. This particular isomer is the only naturally occurring isomer with a half-life more than the age of Universe. The direct decay gammas, or possible $\beta-$ decay from this isomer are unknown, but it can add significantly to our knowledge about the stellar processes if we could estimate its decay rate. The $9^-$ spin of this isomer is due to the spin alignment of proton $\pi 9/2[514]$ and neutron $\nu9/2[624]$ Nilsson orbitals, while the ground state spin $1^+$ has an anti-aligned proton $\pi 7/2 [404]$ and neutron $\nu9/2[624]$ configuration. Due to the large spin difference, a depopulation of this isomer can only occur by photo-excitation into excited states which have a decay branch into the ground states. 

The isomer was first identified in 1955~\cite{Eberhardt1955}, yet the mysterious decay nature of this unique isomer keeps the experimentalists on their toes for half-life measurements. One of the latest half-life measurements has been reported by Lehnert et al.~\cite{Lehnert2017} at HADES underground laboratory in Belgium, which has been compared with the previous measurements from 1955 to 2009~\cite{Eberhardt1955,Bauminger1958,Eberhardt1958,Miller1958,Sakamoto1967,Ardisson1977,Norman1981,Cumming1985,Bissell2006,Hult2006,Hult2006,Hult2009} to establish its half-life, decay, energy and spin. More recently, Nesterenko et al.~\cite{Nesterenko2022} carried out the first direct precise determination of the excitation energy of this natural and low-energy isomer. The latest study by MAJORANA Collaboration~\cite{Arnquist2023} provided an improved half-life limit as $< 0.29 \times {10}^{18}$ $y$ for this isomer. The relevance of this isomer in dark matter search, astrophysics, and development of a $\gamma-$laser makes it a favourite case for researchers. Theoretical understanding of its decay mechanism is also rather limited due to the large-multipolarity involved.  

$^{210}$Bi: There is one more very long lived $9^-$ isomer in $^{210}$Bi $(Z=83, N=127)$ with $3.04 \times {10}^6$ $y$ half-life at 271.31 keV energy. This is a very long-lived alpha-decaying isomer. This nuclide has only one extra proton and neutron than the doubly-closed $^{208}$Pb, providing a simple test for the shell model calculations. The assigned configuration $\pi h_{9/2} \otimes \nu g_{9/2}$ could explain the alpha-decay rate of this isomer~\cite{Mang1964}. 

$^{186}$Re: The $(8^+)$ isomer in $^{186}$Re $(Z=75, N=111)$ has a half-life of $2 \times {10}^5$ $y$ and excitation of 148.2 keV. This isomer is the band-head of a $K=8$ band, decays via 50 keV $E5$ gamma transition to the lower-lying $3^-$ isomeric band-head of a $K=3$ band. The configuration for this isomer has been assigned as $\pi 5/2 [402] \otimes \nu 11/2 [615]$ Nilsson orbitals. Rhenium isotopes $(Z=75)$ belong to the transitional region at $A \approx 190$ where nuclear shape changes from axial deformation to spherical. Theoretical understanding of this axially-deformed isomer is rather limited especially of the decay probabilities and moments. 

\section{$100 \%$ Proton decaying isomers}

A well known mass region to observe $100\%$ proton-emission is the A=150 region, where nuclei having $h_{11/2}$ ground state and $d_{3/2}$, or $s_{1/2}$ excited states are known to decay by proton-emission. The decay of the proton-decaying isomer is thought to occur by a process called proton tunneling, where the proton tunnels through the Coulomb and angular momentum barriers that separates it from the ground state. This process is very rare, and it only occurs in nuclei with very low proton separation energies. 

$^{151}$Lu was the first case of proton-decay from a nuclear ground state observed in 1982~\cite{Hofmann1982}, although the proton-decay of a high-spin isomer in $^{53}$Co has been observed as early as 1970~\cite{Jackson1970}. The ${19/2}^-$ isomer in $^{53}$Co decays to $^{52}$Fe by emitting a proton, though this branch is only 1.5 $\%$. Their potential to provide valuable information about the structure of nuclei at one of the extreme edges of the nuclear landscape makes them very special since no neutron-emitting isomer has been seen until now. Some of the examples are briefly discussed below.

$^{141}$Ho: The ${1/2}^+$ isomer in $^{141}$Ho $(Z=67,N=74)$ with 7.3 $\mu s$ half-life at 66 keV energy. The isomer has been assigned a configuration $\pi 1/2 [411]$ based on the Nilsson scheme~\cite{Ryk1999} and its calculated half-life comes out to be 14.6 $\mu s$~\cite{Barmore2000}. This isomer proton-decays to the ground and first excited $2^+$ state of $^{140}$Dy~\cite{Ryk1999,Sewe2001,Karny2008}.

$^{147}$Tm: The ${3/2}^+$ isomer in $^{147}$Tm $(Z=69,N=78)$ with 0.36 $ms$ half-life at 68 keV energy proton decays to the  ground state of $^{146}$Er~\cite{Toth1993,Sellin1993}. The isomer is the band head of a odd-proton $d_{3/2}$ dominated band while the ground state is the band head of odd-proton $h_{11/2}$ band. 


$^{150}$Lu: An isomer having spin $(1^-,2^-)$ is known in $^{150}$Lu $(Z=71,N=79)$ with 39 $\mu s$ half-life at 22 keV energy. There are seven proton particles above the $Z=64$ sub-shell and three neutron holes below the N=82 shell. In this region, the active single-particle orbitals of protons and neutrons are same: $s_{1/2}$, $d_{3/2}$ and $h_{11/2}$. Neutron and proton couplings result in the ground state and low-lying longer-lived states in odd-odd $^{150}$Lu and eventually lead to two-proton emitting states~\cite{Ginter1999}. In a second experiment performed 20 years ago~\cite{Ginter2003}, the data on proton-emission from the isomeric state was improved, though, no evidence was obtained for fine-structure in the proton emission from its ground state as well as isomeric state. Such cases require special attention at the modern facilities. 


$^{151}$Lu: The ${3/2}^+$ isomer in $^{151}$Lu $(Z=71,N=80)$ with 16 $\mu s$ half-life at 61 keV energy. 
The observation of a second proton transition from this nuclide~\cite{Bingham1999} further motivated a successful search for $^{150}$Lu. The Wentzel–Kramers–Brillouin (WKB) barrier-penetration calculations led to the conclusion that the isomer is a $d_{3/2}$ proton state. The proton-emissions served as sensitive probe of single particle level structure at the drip-line and provided the location of three active single-particle $h_{11/2},$ $d_{3/2},$ and $s_{1/2}$ orbitals in this region.

\section{ $\beta-$decaying isomers}

\begin{figure*}[!htb]
\centering
\includegraphics[width=0.80\textwidth,height=0.65\textwidth]{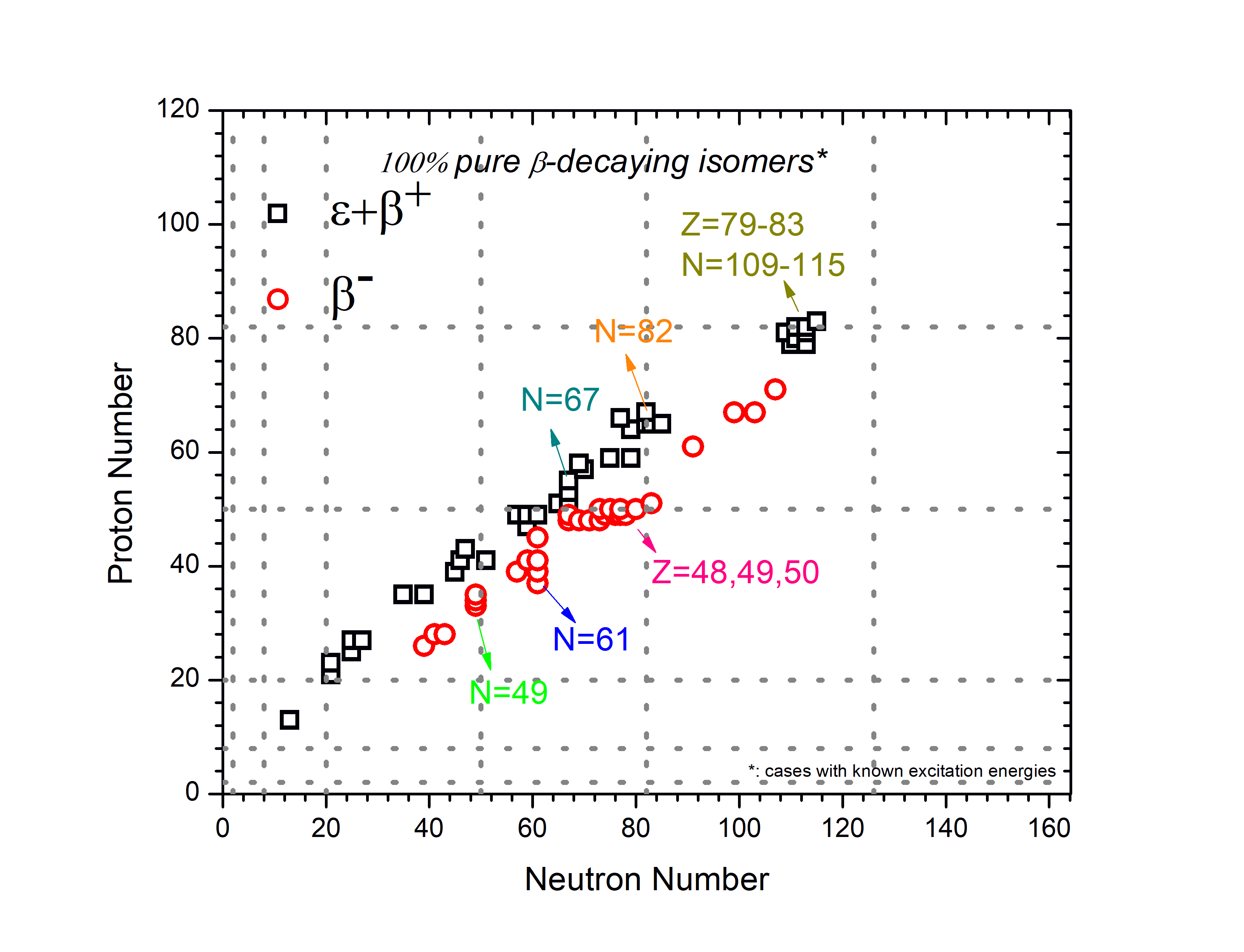}
\caption{\label{fig:betaisomer}(Color online) The occurrence of $100 \%$ pure $\beta-$decaying isomers where isomeric excitation energies are experimentally known; data taken from Atlas of Nuclear Isomers-Second Edition~\cite{Garg2023}. The complete list of these cases is available in the book,``Nuclear Isomers-A Primer"~\cite{Book}.  } 
\end{figure*}

Isomers which entirely decay by beta decay processes may be of special interest in nuclear astrophysics. It is known that beta decay is a slow process and an excited state will always try to find a gamma decay path which is much faster than beta decay. In case of $100 \%$ beta decay, such a path is suppressed and, therefore, only beta decay occurs. We plot the $100 \%$ $\beta-$decaying isomers in Fig.~\ref{fig:betaisomer}, particularly the cases with known excitation energies. Those beta decaying isomers which have even a small branching by gamma/other decay mode, or, their excitation energy is not known, have not been included; such cases are quite a few. Most of the $\beta-$decaying isomers have very low excitation energy making them challenging for experimentalists  to separately distinguish them from their ground states. The modern radioactive ion beam facilities provide a way to approach them more carefully especially in exotic nuclei such as a new 6.6 milliseconds $\beta$-decaying isomer identification in the neutron-rich $^{36}$Al~\cite{Lubna2023}. 

Many $\beta-$decaying isomers have half-lives longer than their respective ground states. For example, $^{42}$Sc has a ground state of $0^+$ with 687.7 ms half-life while the $\beta^+$-decaying isomer has a half-life of 61.7 sec. Similar is the situation in $^{44}$V, $^{50}$Mn, $^{54}$Co, $^{70}$Br, $^{74}$Br, $^{106}$Ag, $^{116}$Sb, $^{118}$Sb, $^{122}$Cs, $^{138}$Pr among other $\beta^+$ isomers. 

Similar situation is also encountered in $\beta^-$ isomers. One of most long-lived cases is the odd-odd $^{166}$Ho having a $100\%$ $\beta^-$decaying isomer with 1132.6 years half-life at 5.9 keV energy while its ground state has a half-life of 26.824 hours only. It was first identified accidentally in a thermal neutron $(n, \gamma)$ reaction measurement trying to populate $^{167}$Ho but ended up finding the longer-lived state of $^{166}$Ho~\cite{Butement1951}. This isomer with 1132 years of half-life is capable of slowing down the stellar beta-decay rates~\cite{Misch2021}. It is increasingly being recognised that isomers may provide alternate reaction paths; some of them are also important in nuclear fission and astrophysical reactions~\cite{Lotay2022,Colonna2020}. 

Isomers crucial in astrophysical environments are now referred to as astromers~\cite{Misch2021}; a separate article on this topic is included in this special issue. $^{26}$Al isomer is one of the famous astromers with a $N=Z$, odd-odd configuration. There are few other $N=Z$ systems also having $\beta^+$ decaying isomers in lighter-mass odd nuclei such as $^{42}$Sc, $^{50}$Mn, $^{54}$Co, and $^{70}$Br. 

\textbf{Isomers around Z=50}: As an example, let us focus on the region around $Z=50$. There is a systematic of ${11/2}^-$ $\beta^-$decaying isomers in $^{115,117,119,121}$Cd isotopes dominated by the unique parity $h_{11/2}$ neutron orbital. On the other hand, in $^{123,125,127}$ Sn isotopes, the ${11/2}^-$ state becomes the ground state while the ${3/2}^+$ state transforms into a $\beta^-$decaying isomer. Evolution of isomers is, therefore, turning out to be of great use in understanding the single-particle energies at the Fermi surface and their evolution with particle number near the drip-lines. There is also a chain of $2^+$, $\beta^+$decaying isomers in odd-odd $^{106,108,110}$In $(Z=49)$ isotopes with half-lives of 5.2, 39.6 and 69.1 minutes, respectively. The isomeric excitation energies in all three cases are very low, of the order of 30-60 keV. All three have a $7^+$ ground state. There are $8^-$ $\beta^+$decaying isomers in odd-odd $^{116,118}$Sb $(Z=51)$ isotopes with half-lives of 60.3 min and 5 hours, respectively. Both these cases are longer-lived than their respective ground states with half-lives of 15.8 min and 3.6 min.     

Odd-odd nuclei again dominate the scene in $\beta-$decaying isomers. Most theoretical frameworks face challenges in describing odd-odd nuclei, let alone isomers. As an example of a recent study, three beta decaying isomeric states in odd-odd $^{128}$In and $^{130}$In have been studied with the JYFLTRAP Penning trap at the IGISOL facility by Nesterenko \textit{et al.} ~\cite{Nesterenko2020}. A new beta-decaying high-spin isomer has been discovered in $^{128}$In at 1797.6 (20) keV, which is suggested to be a ${16}^+$ spin-trap using shell-model calculations. For the first time, the lowest-lying (${10}^-$) isomeric state at 58.6(82) keV was resolved in $^{130}$In using the phase-imaging ion cyclotron resonance technique. Such precise measurements on the energies of the excited states are quite crucial for improving the shell-model effective interactions near $^{132}$Sn.    

\section{Isomer usage in dark matter search}

Direct detection experiments face significant challenges in constraining inelastic dark matter and strongly interacting dark matter search because both require the scattering event to transfer energy from the nucleus into the dark matter. A new and interesting possibility has been highlighted to use nuclear isomers as dark matter accelerators by Pospelov, Rajendran and Ramani~\cite{Pospelov2020}. The collisional deexcitation of isomers by dark matter can have a momentum exchange with the nucleus in its meta-stable state. This  
can be observed either by searching for the direct effects of the decaying isomer, or through the decay of excited dark matter states in a conventional dark matter detector setup. The potential candidates for this could be the naturally occurring isomer in $^{180}$Ta, 31 years isomer in $^{178}$Hf, 2.55 min isomer in $^{137}$Ba
produced from Cesium in nuclear waste, or 160 days isomer in $^{177}$Lu from medical waste. This is a futuristic goal and will be a milestone in modern physics if it succeeds.    

\section{Conclusions}

To conclude, nuclear isomers are becoming important not only in fundamental nuclear physics and astrophysics but also in industrial applications such as medicine, energy, and nuclear clocks etc. Due to which, isomer physics is currently trending around the world. In this article, we have highlighted, with specific examples, those isomers which carry some special property or, characteristic, making them unique among their kin. We have thus discussed isomers having high energy, high multipolarity decay, high spin, large K value etc. We also bring out the specialities of isomers with extremely low energy, proton-decay, and beta-decay. New terms like Astromers, ELE isomers, etc are being coined to designate special types of isomers.  

High-energy isomers usually correlate with the large shell gaps in spherical as well as deformed nuclei. These isomers are very useful in unravelling the evolution of changing magic numbers especially in exotic and neutron/proton-rich nuclei. High multipolarity isomers generally arise due to the coupling of intruder orbitals with their neighboring low-j orbitals and belong to the spin isomers, if the isomeric transition is about $100 \%$.

ELE isomers emerge as a novel class of isomers primarily controlled by extremely low-energy transitions without involving hindrances arising from the large change in angular momentum or projection of angular momentum. These isomers are crucial in potential isomer applications and are useful probes for the atomic-nuclear interface through the emission of low-energy gamma rays.

Very high-spin isomers generally imply the involvement of many nucleon-pair breakups. High-spin and high-energy isomers mostly belong to the region around $^{208}$Pb and $^{146}$Dy where the nuclei exhibit limited deformation but feature the presence of single-particle high-j orbitals alongside octupole vibration at higher energies resulting from core-excitations. While investigations into such cases extend to the trans-lead region, data in the $^{146}$Dy region is relatively scarce. 

Odd-odd nuclei due to complex proton-neutron couplings near the Fermi surface may lead to high-spin isomers at low energies such as ${11}^-$, 241 $y$ isomer in $^{192}$Ir. These isomers are often very long-lived due to involved large spin changes and may also exhibit large multipolarity transitions.

Proton decaying isomers, as implied by their name, are located in the proximity of the proton-drip line which represents one of the extremes in nuclear chart. Such isomers may be potential candidates for applications in proton therapy.
$\beta-$decaying isomers are also situated at the extremes as well as near the $\beta-$stability line of normal nuclides. Undoubtedly, many of them intersect with the pathways of stellar nuclear reactions, exerting an influence on the elemental abundances, a topic of current interest.

While we do possess a qualitative knowledge of isomer hindrance mechanisms and configurations, a comprehensive understanding of isomeric decay mechanisms, particularly in the deformed region, remains elusive. The decay probabilities in isomers are susceptible to small changes in the wave function overlaps between the initial and final states. This complexity further increases when multiple excitations are involved. A consistent understanding of isomeric decays and moments pose a good challenge to the modern nuclear theories.

\noindent \textbf{Acknowledgements:}
BM gratefully acknowledges the financial support from the Croatian Science Foundation and the \'Ecole Polytechnique F\'ed\'erale de Lausanne, under the project TTP-2018-07-3554 ``Exotic Nuclear Structure and Dynamics'', with funds of the Croatian-Swiss Research Programme.

\noindent \textbf{Data Availability Statement:}  No Data associated in the manuscript.

\end{document}